\documentclass[aip,jcp,reprint,superscriptaddress]{revtex4-1}
\usepackage{graphicx}
\usepackage{comment}
\usepackage{amsmath}
\usepackage{wasysym}
\usepackage[usenames, dvipsnames]{color}
\begin{document}


\title{Nonequilibrium quantum solvation with a time-dependent Onsager cavity}

\author{H. Kirchberg}
\affiliation{I. Institut f\"ur Theoretische Physik, Universit\"at Hamburg, Jungiusstra\ss{}e 9, 20355 Hamburg, Germany}
\author{P. Nalbach}
\affiliation{Westf\"alische Hochschule, M\"unsterstr. 265, 46397 Bocholt, Germany}
\author{M. Thorwart}
\affiliation{I. Institut f\"ur Theoretische Physik, Universit\"at Hamburg, Jungiusstra\ss{}e 9, 20355 Hamburg, Germany}

\date{\today}

\begin{abstract}
We formulate a theory of nonequilibrium quantum solvation in which parameters of the solvent are explicitly depending on time. We assume in a simplest approach a spherical molecular Onsager cavity with a time-dependent radius. We analyze the relaxation properties of a test molecular point dipole in a dielectric solvent and consider two cases: (i) a shrinking Onsager sphere, and, (ii) a breathing Onsager sphere.  
Due to the time-dependent solvent, the frequency-dependent response function of the dipole becomes time-dependent. For a shrinking Onsager sphere, the dipole relaxation is in general enhanced. This is reflected in a temporally increasing line width of the absorptive part of the response. Furthermore, the effective frequency-dependent response function shows two peaks in the absorptive part which are symmetrically shifted around the eigenfrequency. In contrast, a breathing sphere reduces damping as compared to the static sphere. Interestingly, we find a non-monotonous dependence of the relaxation rate on the breathing rate and a resonant suppression of damping when both rates are comparable. Moreover, the line width of the absorptive part of the response function is strongly reduced for times when the breathing sphere reaches its maximal extension. 
\end{abstract}


\maketitle

\section{Introduction}
\label{secIntro}
Electronic spectroscopy of photoexcited molecules dissolved in a polar medium measures the response of molecular electronic states to applied time-dependent electric fields and, by this, also reveals the interplay of the dissolved molecule (the solute) and the solvent \cite{Review,MayBook}. Since the early days of quantum mechanics, it has been established that the solvent induces pronounced features in the molecular response. Prominent examples are the broadening of spectral lines, and the Stokes shift between absorption and emission lines which both reflect the microscopic influence of the solvent \cite{MayBook,Review,NitzanBook,bot1973}. 

Different approaches to describe solvation exist,  see Ref.\  \onlinecite{Review} for a review. 
One particularly simple picture of quantum solvation describes the solvent as a polarizable continuous, homogeneous and isotropic dielectric medium with a complex dielectric function $\varepsilon(\omega)$. The solute  is assumed to possess a certain charge distribution. When placed in the solvent, it polarizes the  solvent, which itself produces a reaction electric field for the charge distribution of the solute, leading to a solute-solvent interaction. 

The earliest polarizable solvent continuum model was developed by Born in 1920 \cite{bor1920} who assumed a solute with a spherical charge distribution in the center of an effective molecular cavity with an unknown time-independent radius. This molecular cage itself is placed in a classical dielectric continuum. The solute-solvent interaction is determined by  the net electric charge of the solute. Later, Kirkwood \cite{kir1935} and Onsager \cite{ons1936} extended this model to include the second and higher-order multipole terms of the electrostatic solute potential. The solute is described as a point dipole located at the center of a spherical (Kirkwood) or an ellipsoidal  (Onsager) molecular cavity which, again, is static and whose unknown parameters are empirical and used as fit parameters.  Inside the molecular cavity, vacuum is assumed. The model gives then rise to a  dipole-dipole interaction between the central molecule and the solvent. This technique captures properties of the solute in its ground state under conditions of thermal equilibrium.

Nonequilibrium processes and time-dependent solvation are important when external time-dependent electric fields are applied and photoexcitation occurs \cite{Review,MayBook,NitzanBook}.  Usually, the applied electric fields only couple to the solute so that the solvent remains close to its thermal equilibrium.  A minimal model of the solute then encompasses two states, a ground state and an excited state. Time-dependent relaxation and dephasing  in the solute in the presence of the solvent carry information on the solute-solvent interaction. The solvent forms a fluctuating quantum dissipative harmonic environment at thermal equilibrium \cite{wei2012,leg1987,MayBook,NitzanBook}. The equilibrium statistics of the environmental fluctuations determines relaxation and dephasing times of the solute and leads to spectral line broadening. The bath characteristics can be encoded in the bath spectral density 
\cite{wei2012} $J(\omega)$. For the Onsager model, $J(\omega) \propto$ Im $\varepsilon(\omega)$. The basis  is the assumption that time-dependent changes in the  electronic configuration of the solute are not too large such that the back action of the solvent polarization on the solute can be calculated in terms of linear response theory. Then, the fluctuation-dissipation theorem holds, the fluctuations are thermal, and a bath spectral density exists in the established form \cite{wei2012}. 

The description of dynamical quantum solvation in terms of the Kirkwood-Onsager model has been refined to include the effect of a hydration shell of tightly bound solvent molecules which are modelled by a shell of a continuous dielectric medium \cite{mck2005,nal2014}. The total dielectric function determines a more complicated spectral density with a, in general, non-Markovian relaxtaion dynamics \cite{mck2005,nal2014}. A thicker hydration shell provides an enhanced source of fluctuations such that the lifetime of the excited solute state is noticeably reduced. Yet, the solvent is still close to its thermal equilibrium. 

An important parameter of the equilibrium quantum solvation theory is the radius $a_0$ of the static molecular cavity. It enters in the spectral density as $J(\omega)\propto 1/a_0^3$ in the case of a sphere. A recent study \cite{bre2014} of a sequence of electronic transitions between a low-spin and a high-spin electronic configuration of aqueous Fe-II complexes has provided possible hints of 
 fast dynamical changes in the first hydration shell which accompany the electronic transitions in the central molecule.  In the experiment, an increase of the lengths of the bonds between the central Fe-atom and the nearest neighbor atoms in the complex was identified as being induced by the electronic transitions. The experimental data have been interpreted such that the transition from the low-spin to the high-spin configuration due to photoexcitation is accompanied by a structural change of the hydration layer: On average two shell water molecules are expelled from the hydration shell into the bulk solvent. The distribution of the water molecules in the shell is modified  during the transition  influencing the electronic relaxation. This experiment is consistent with earlier theoretical predictions \cite{dak2010,pen2012} obtained on the basis of ab initio molecular dynamics simulations. 

The time-dependent rearrangement of the molecular cavity has also been revealed for aqueous iodide by a joint 
experimental and theoretical analysis \cite{pen2013}. Static and time-resolved X-ray absorption spectroscopy have disclosed a 
transition from hydrophilic to hydrophobic solvation after an electron has been removed from the iodide. This goes along with a rapid expansion of the cage within $200$-$300$ fs. The formation of a new shell of water molecules around iodine is more involved and occurs within $3$-$4$ ps. 
A single water molecule is pulled
towards the bulk of the solvent \cite{pen2013}, which has been confirmed by quantum chemical and molecular dynamics simulations. Alternative approaches formulate  equations of motion for the solvent polarization charges  and couple them to a real-time time-dependent density functional theory  \cite{Corni1} or a 
time-dependent configuration interaction description \cite{Corni2}. 
Another example is the structural dynamics of photoexcited of  [Co(terpy)$_2$]$^{2+}$ in aqueous solution investigated with ultrafast x-ray diffuse scattering \cite{bia2016}. Accompanying density functional theory calculations showed  that the photoexcitation leads to elongation of the Co-N bonds.  Finally, we note that externally controlled time-dependent configurational changes have also been shown to impact the charge transfer in metal donor-bridge-acceptor complexes \cite{del2014,lin2009,lyn2012}.

Up to present, a generalization of the Kirkwood-Onsager continuum equilibrium theory of quantum solvation to explicitly include time-dependent solvent properties has not been established. In this work, we formulate a theory of nonequilibrium quantum solvation in which parameters of the solvent become explicitly time-dependent. We assume in a simplest approach a spherical molecular Onsager cavity with a time-dependent radius. The precise functional form of the time-dependence is imposed from outside and we consider two cases: (i) a shrinking Onsager sphere, and, (ii) a breathing Onsager sphere with a transient expansion followed by a shrinking to the original size \cite{bia2016}. In particular, we are interested in how the relaxation properties of an excited electronic state of the solute are influenced by the time-dependent solvent cage and how its frequency-dependent response characteristics is. We consider a molecular point dipole dissolved in water. Since all 
parameters are kept explicit, other dielectric solvents can readily be addressed as well. In general, the frequency-dependent response functions are non-stationary and explicitly depend on time. For a shrinking Onsager sphere, the fluctuating solvent comes closer to the solute such that relaxation in the solute becomes faster and damping is overall stronger. This is reflected in a  frequency-dependent response function which is  temporally enhanced up to $40 \%$ and which drops down with increasing time. For a slower shrinking, the magnitude of the absorptive part of the response is enhanced for longer times while its width is reduced. Correspondingly, the relaxation rate decreases as well for slower shrinking. A breathing sphere generates reduced damping as compared to the static sphere. The line width of the absorptive part of the response is therefore depleted at times when the breathing radius reaches its maximum. Accordingly, we find a non-monotonous dependence of the relaxation rate on the breathing rate and a resonant suppression of damping when both rates are comparable.

From the point of view of the theory of dissipative quantum systems, a time-dependent hydration shell amounts to a time-dependent system-bath coupling and to a break-down of the stationarity of the environmental fluctuations. 
This calls for a generalization of the formalism to nonequilibrium baths. In particular, a standard form of the spectral density connected to a time-translationally invariant susceptibility and a damping kernel no longer exists \cite{wei2012}. This is different to the previously considered case of an externally driven harmonic bath in which an external time-dependent field drives the harmonic bath oscillators \cite{drivenbath1,drivenbath2}. By this, a net external force on the central systems is composed by the bath, but it still can be described by a spectral density. 

The paper is structured as follows: In Sec.\ \ref{secModel}, we describe the model of nonequilibrium quantum solvation with a time-dependent cavity radius and derive the governing equation of motion for the polarization of a test molecule in the form of a point dipole as the solute. Then, we carefully select the model parameters in Sec.\ \ref{sec.modelparam}. The case of a shrinking Onsager sphere is presented in Sec.\ \ref{secShrink}, while a breathing sphere is analyzed in Sec.\ \ref{secBreath}, before we conclude the work.

\section{Model and equation of motion} \label{secModel}

We consider a solute molecule in the form of an electric dipole which is exposed to the reaction electric field  generated by the back action of the polarized solvent. The electrons are bound to the positively charged nuclei and can be brought out of their equilibrium position by the fluctuating reaction field. In the simplest form of a homogeneous, isotropic model, the effective potential for the electron (with a charge $e$) near its equilibrium position is assumed harmonic with the characteristic frequency $\omega_0$, 
and the fluctuations induce damping of its time-dependent displacement $q(t)$ \cite{jac1999,weg2004}. The induced dipole moment of the solute is then given by $\mu (t)= e\, q(t)$. The dipole is placed inside a spherical cavity whose  radius $a(t)$ is now assumed to be explicitly time-dependent. We assume that the time-dependence is created by some external mechanism, for instance, upon photoexcitation of the solute and a subsequent reconfiguration of the solute or the solvent (see the discussion in Sec.\ \ref{secIntro} for specific examples). The spherical cage is surrounded by a dielectric medium with a given frequency-dependent complex permittivity $\varepsilon(\omega)=\varepsilon'(\omega)+i\varepsilon''(\omega)$. 

The time-varying dipole moment $\mu(t)$ induces an electric field in the surrounding dielectric medium which thereby gets polarized. This solvent polarization generates an electric field component (the Onsager reaction field $R(t)$ \cite{bot1973,ons1936}) at the position of the central solute dipole and alters its electrostatic energy due to dipole-dipole interaction. Since the solvent is assumed as a continuous medium, the reaction field is a fluctuating quantity, which consequently generates damping for the motion of the central dipole. 

To calculate the Onsager reaction field by means of electrostatics for a time-dependent radius, we 
assume for simplicity that the radius $a(t)=a_0 +a_1(t)$ is composed of a time-independent part $a_0=$const.\ and only changes weakly in a small range by the magnitude $a_1(t)$ over time. This allows us to neglect any charge current density and to use the much simpler laws of electrostatics \cite{bot1973,ons1936}. Exploiting the spherical symmetry, one can find the electrostatic potential in the region outside the Onsager sphere in the dielectric medium ($r>a(t)$) as 
\begin{align}
\phi_A(t)=\frac{\mu_{\rm ind}(t)\cos(\theta)}{r^2}\, ,
\end{align}
where $\mu_{\rm ind}(t)$ is the dipole moment induced in the surrounding dielectric medium which needs to be determined and $\theta$ the polar angle relative to the direction of the dipole moment $\mu$. Inside the sphere ($r<a(t)$), the electrostatic potential is determined to be 
\begin{align}
\phi_I(t)=R(t)r\cos(\theta)+\frac{\mu(t)\cos(\theta)}{r^2}\, ,
\end{align}
where the first term gives the potential of the reaction field caused in response of the dielectric medium to the dipole moment $\mu(t)$, while the second term represents the potential field of the dipole moment itself. $\mu_{\rm ind}(t)$ and $R(t)$ are determined by the boundary conditions at the momentary point $r=a(t)$, namely, the requirement of continuity of the overall potential [$\phi_I=\phi_A$] and of the continuity of the normal component of the electric displacement field [$\partial\phi_I/\partial n = \varepsilon \partial\phi_A/\partial \phi n$]. All contributions of the magnetic field induced by the displacement current at the boundary of the slowly changing sphere are negligible. With this, the reaction field is found to be 
\begin{align}\label{reactionfield}
R(t)=\frac{1}{a(t)^3}\int_0^t dt' \chi(t-t')\mu(t') \, .
\end{align}
The dipole moments of solvent molecules do not adjust instantaneously to the central dipole moment but rather lag behind the changing dipole in time. In frequency space, one finds that the response function is related to the permittivity by 
\begin{equation}\label{chi}
\chi(\omega) = \frac{\varepsilon(\omega)-1}{2\varepsilon(\omega)+1} \, .
\end{equation} 
Since we consider a homogeneous isotropic continuous solvent, we assume the Debye approximation \cite{NitzanBook} for the permittivity, i.e., 
\begin{equation}
 \varepsilon_D (\omega) = \varepsilon_\infty + \frac{\varepsilon_s -\varepsilon_\infty}{1-i\omega\tau_D} \, ,
\end{equation}
with the zero frequency static dielectric constant $\varepsilon_s$ and the dielectric
constant at asymptotic frequencies $\varepsilon_\infty$. The Debye relaxation time is
denoted as $\tau_D$. Transforming back to the time domain, one finds the response function 
\begin{align}
\chi(t)=\chi_D e^{-\omega_D t}\Theta(t),
\label{6}
\end{align}
where we have defined $\omega_D=(2\varepsilon_s+1)/(3 \tau_D)$ and $\chi_D= \frac{2(\varepsilon_s-1)}{3\tau_D}$.
Moreover, we define the Laplace transform by $f(z)=i\int_0^\infty dt \, e^{izt} f(t)$, where $f(z)$ is analytic for $\rm{Im}(z)>0$ and $f(t\to \infty) < \infty$. This yields the response function in the Laplace space according to
\begin{equation}
\chi (z) = - \frac{\chi_D}{z+i \omega_D} \, .
\end{equation}

We now understand the reaction field $\langle R(t) \rangle$ as force operator whose expectation value gives rise to a time-dependent expectation value for the external force $e\langle R(t) \rangle$ leading to a linear equation of motion for the expectation value $\langle q\rangle$ of the displacement of the electron with mass $m$, i.e., 
\begin{equation}\label{1}
m\langle\ddot{q} (t)\rangle+m\omega_0^2\langle q(t) \rangle =\frac{e^2}{a(t)^3}\int_0^t dt' \chi(t-t')\langle q(t') \rangle \, , 
\end{equation}
where $\langle q(t) \rangle=0$ for $t<0$ has been assumed. In the frequency domain, the response function $\chi(\omega)=\chi'(\omega)+i\chi''(\omega)$ can be separated into the real and imaginary part, where the real part $\chi'(\omega)$ leads to a shift of the potential $V(q)$ and the imaginary part $\chi''(\omega)$ induces damping of the displacement motion. We note that for a static Onsager radius $a(t)=a_0$, the equation of motion can be brought into the usual form involving a velocity-dependent friction force term which is non-local in time. The damping kernel can be associated to the spectral density $J(\omega)$ of an equilibrium environment (see Supplementary Material). For the nonequilibrium solvent with a time-dependent Onsager radius, this is no longer possible, since the fluctuations of the nonequilibrium bath are no longer time-translational invariant (or, stationary). 

In the following, we shall use the equation of motion in the form of Eq.\ (\ref{1}).  We first consider a nonadiabatic shrinking of the Onsager sphere in the section \ref{secShrink}, before we treat a breathing sphere which first expands and then shrinks back to the initial configuration.  

\section{Model parameters}
\label{sec.modelparam}
Throughout this work, we study a test molecule with an eigenfrequency $\omega_0=1.183 \times 10^{15}$ Hz, which sets the natural system time scale of the dipole dynamics comparable to these ones of molecules and atoms in the ultraviolet regime. Furthermore, our choice of  $\omega_0$ is restricted to the regime above a minimal magnitude for stability reasons within the model (see Sec.\ \ref{subsecTimevol}). The relevant mass is given by the electron mass $m=9.11 \times 10^{-28}$ g.  Then, the natural length scale of the displacements $q(t)$ is given by the 
oscillator length $q_d=\sqrt{\hbar/(m\omega_0)}=3.1$ \AA$=3.1 \times 10^{-8}$cm (note that we work in cgs units in which the electron charge is $e=3\times 1.602 \times 10^{-10}$esu). 

For the solvent, we use the values of bulk water (at $20^\circ\text{C}$) according to $\varepsilon_s=78.3$, 
$\tau_D= 8.2$ ps of the dipolar solvent and $\omega_D=6.4$ THz $=5.4 \times 10^{-3} \omega_0$. Moreover, $\chi_D=6.3$ THz. For simplicity, we set  $\varepsilon_\infty=1$ in the high frequency limit, while other values ($\varepsilon_\infty>1$) for this limit lead to small constant shifts of the potential (see Supplementary Material for details).The scale $a_0$ of the radius of the static Onsager sphere is assumed to be $a_0 = 6 \times 10^{-8}$cm. Hence, $\omega_0 \gg \omega_D, \chi_D$, and under these conditions, the scale of the right-hand side of Eq.\ ({\ref{1}}) becomes $\sqrt{e^2/(m a_0^3)}=1.1 \times 10^{15}$ Hz $\simeq \omega_0$. 

\section{Shrinking Onsager sphere}
\label{secShrink}
First, we consider the relaxation dynamics of the dipole in a shrinking Onsager sphere. We assume that the radius varies with time from the larger radius $a_0+a_1$ to the smaller one $a_0$ according to 
\begin{align}
a(t)=a_0+a_1(t)=a_0+a_1 e^{-\alpha t} \Theta(t). \label{4}
\end{align}
$\alpha$ denotes the shrinking rate. 

From now on, we write all expectation values as $\langle q(t) \rangle \equiv q(t)$ and $\langle R(t) \rangle \equiv R(t)$ for better readability.

\subsection{Time evolution of the dipole moment}
\label{subsecTimevol}
To calculate the expectation value of the dipole moment $\mu(t)$, we need to solve Eq.\ (\ref{1}) which is a linear integro-differential equation. This is not possible in general in an analytical manner. To proceed, we employ the Laplace transform and need to use an approximation for the prefactor $1/a(t)^3$. In doing so, we expand the prefactor in a Taylor series up to first order in the time-dependent part. This is valid when the overall magnitude $a_1$ by which the radius changes over time is much smaller than the static radius $a_0$, which is usually the case in a concrete physical situation. In fact, the molecule is typically much larger than the change of its size induced by photoexcitation.

Hence, we have for $a_1 \ll a_0$ that 
\begin{align}
\frac{1}{a(t)^3}=\frac{1}{(a_0+a_1e^{-\alpha t})^3}\simeq\frac{1}{a_0^3}\left[1-\frac{3a_1}{a_0}e^{-\alpha t}
\right]\, . 
\label{2b} 
\end{align} 
For convenience, we define the rescaled linear susceptibilities for the case of fast shrinking according to 
\begin{align}
\label{37}
\chi_0 (t) & =  \frac{e^2}{ma_0^3} \chi(t) \\ \notag
\chi_1 (t) & = \frac{3a_1}{a_0}\chi_0(t) \, ,
\end{align}
here,  $\chi_0 (t)$ only depends on the static radius $a_0$ while $\chi_1 (t)$ is of first order in $a_1/a_0$. 

In first order in the ratio $a_1/a_0$, we obtain the Laplace transform of Eq.\ (\ref{1}):
\begin{align}
-z+[\omega_0^2-z^2+i\chi_0(z)]q(z) =i\chi_1(z+i\alpha)q(z+i\alpha)\, .
\label{14}
\end{align}
The solution immediately follows as 
\begin{align}
\label{30d}
q(z)&= q_0(z) + q_1(z) \\ \notag
&= \frac{z q(t=0)}{\omega_0^2-z^2+i\chi_0(z)} 
+\frac{i\chi_1(z+i\alpha)q_0(z+i\alpha)}{\omega_0^2-z^2+i\chi_0(z)}\, .
\end{align}
Here, $q(t=0)$ denotes the initial expectation value of the displacement at $t=0$ and we have assumed that the initial expectation value of the velocity 
$\dot{q}(t=0)=0$. For specific calculations below, we set $q(t=0)=q_d$. Moreover, we have defined $q_0(z)$ which includes the contribution from the static smaller sphere with the final radius $a_0$, while all contributions from the time-dependent radius are contained in $q_1(z)$. Since we need to make sure that the second term $q_1(z)$ only contains contributions up to first order in $a_1/a_0$ (which is realized by the prefactor $\chi_1(z+i\alpha)$), we need to replace $q(z+i\alpha)$ by $q_0(z+i\alpha)$ here. 

Thus, we first calculate $q_0(z)$ for the static case by setting $a_1=0$. Then, we calculate the first-order term $q_1(z)$ which reflects the influence of the time-dependent radius $a_1(t)$. Finally, we transform the results back to the time domain to obtain the overall time-dependent expectation value of the displacement $q(t)=q_0(t)+q_1(t)$ (see Supplementary Material for details).

To obtain a stable solution of $q_0(t)$ as Laplace back transform of $q_0(z)$, we have to restrict the frequency domain such that $\omega_0>\omega_{0, \rm{min}}$ (see Supplementary Material for details). If we choose a smaller eigenfrequency, we will obtain an additional solution of $q_0(t)$ which infinitely grows in time. The central molecular dipole moment would then be completely distorted beyond the harmonic limit, and a real molecule, based on the Onsager model, would be deformed in terms of its extensional and electronic properties which is unphysical. 

The resulting zero order $q_0(t)$ is determined to be 
\begin{align}
\label{10}
q_0(t)&=e^{-\Gamma_0 t}q(t=0)\cos[\omega_0 t],
\end{align}
with the damping time constant 
\begin{equation}\label{10a}
\Gamma_0=\frac{1}{2}(u+v)+\frac{\omega_D}{3} \, ,
\end{equation}
with $u=(-r/2+\sqrt{\Delta})^{1/3}$ and $v=(-r/2-\sqrt{\Delta})^{1/3}$ and 
\begin{equation}\label{10b}
\Delta=\left( \frac{r}{2}\right)^2 + \left( \frac{p}{3}\right)^3 \, ,
\end{equation}
and
\begin{eqnarray}\label{10c}
p&=&-\frac{1}{3}\omega_D^2 + \omega_0^2 \nonumber \, , \\
r&=&\frac{2}{27}\omega_D^3 + \frac{2}{3}\omega_0^2 \omega_D+ \frac{e^2}{ma_0^3}\chi_D \, .
\end{eqnarray}
For a fixed value of $a_0$, it follows that $\Gamma_0\simeq \omega_D/3$. 
%
It is important to note that we have used here the model parameters of water and the relations between them as stated in Sec.\ \ref{sec.modelparam}. Only then, the damping strength is given by Eq.\ (\ref{10}), the oscillation frequency is given by $\omega_0$, and other components (such as that coming from the sine) are negligible. For the general case, the expressions get more involved and we refer to the Supplemental Material. Moreover, we note that for a fixed parameter combination and without the neglect of the small terms, $\Gamma_0\propto 1/a_0^3$, i.e. an increased radius results in a smaller damping \cite{nal2014}.

\begin{figure}[t] 
 \begin{center}    
   \includegraphics[width=0.3\textwidth, angle=-90]{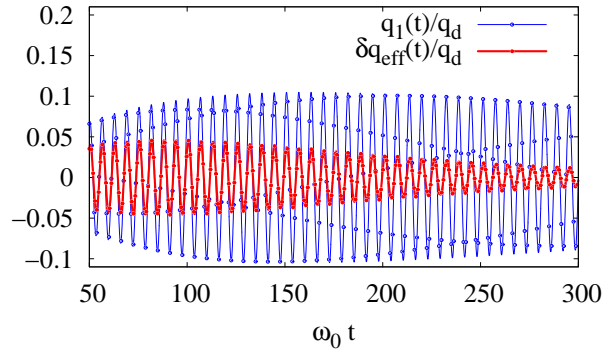}
  \caption{\label{fig1} Time-dependent first-order contribution $q_1(t)$ for $a_1/a_0=0.1$ and for the shrinking rate $\alpha=5\Gamma_0$ (blue) for the parameters given in Sec.\ \ref{sec.modelparam}. The red curve shows the estimated effective correction $\delta q_{\rm{eff}}(t)$ obtained by substituting $a_0$ by the full time-dependent radius $a(t)$ in the expression for $q_0(t)$ together with Eqs.\ (\ref{10})-(\ref{10c}). 
  }
\end{center} 
\end{figure}  

\begin{figure}[t] 
\centering 
   \includegraphics[width=0.3\textwidth, angle=-90]{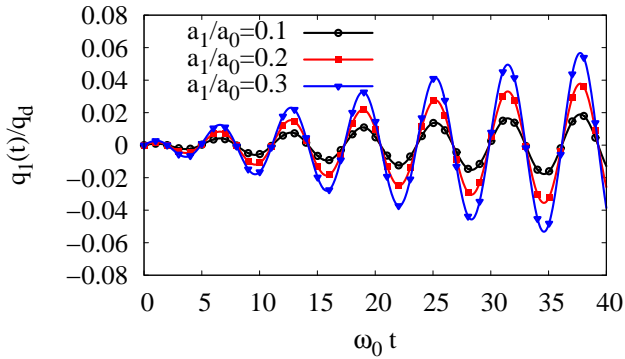}
  \caption{\label{fig2} Time-dependent first-order contribution $q_1(t)$ for short times for different values of $a_1/a_0=0.1$ (black), $0.2$ (red), and $0.3$ (blue) for $\alpha=5\Gamma_0$.}
\end{figure}

With this result at hand, we investigate next the impact of the temporally shrinking radius $a_1(t)$ of the Onsager sphere through $q_1(t)$. The first-order term $q_1(t)\sim a_1/a_0$ is obtained from the inverse Laplace transform of $q_1(z)$ in Eq.\ (\ref{30d}). The expression is somewhat involved and we refer to the Supplementary Material for details. The result is shown in Fig.\ \ref{fig1} for $a_1/a_0=0.1$ and for a rather fast shrinking with $\alpha=5\Gamma_0 \simeq 9 \times 10^{-3}\omega_0$. The additional term is in phase with $q_0(t)$ and has the same frequency as $q_0(t)$. Its amplitudes first grow, reach a maximum and finally decay to zero. Its behavior at short times is shown in Fig.\ \ref{fig2} for different values of $a_1/a_0$. For the chosen shrinking rate $\alpha=5\Gamma_0$ the maximum magnitude of $q_1(t)$ is reached around $t\approx 150 \, \omega_0^{-1}$. After that, the central dipole experiences the shrinking sphere and thus an enhanced damping, such that the amplitude decreases. The maximum amplitude occurs on a time scale $\alpha^{-1} \simeq 120 \, \omega_0^{-1}$. For an increased shrinking rate, the maximum magnitude of the additional $q_1(t)$ is reached earlier as the sphere shrinks faster.  A larger magnitude 
of $a_1$ results in a bigger initial radius $a(t=0)=a_0+a_1$ of the Onsager sphere where the fluctuating solvent molecules are located further away from the dipole. This gives rise to a more rapid initial increase in $q_1(t)$, see Fig.\ \ref{fig2}.  During the shrinking,  the fluctuating solvent molecules come closer and their impact increases which happens on the characteristic timescale $\alpha^{-1}$, before going to zero at infinite time when the final radius $a_0$ is reached.

As the magnitude $a_1$ by which the radius changes is much smaller than the static radius $a_0$, one may approximate the time-dependent $q(t)$ by an effective $q_{\rm eff}(t)$ which is obtained from $q_0(t)$ in Eq.\ (\ref{10}) with the constant decay rate $\Gamma_0$ being replaced by a time-dependent decay rate $\Gamma(t)\propto 1/(a(t))^3$. The latter is obtained by replacing $a_0$ in Eq.\ (\ref{10c}) by $a(t)$. The resulting $\delta q_{\rm{eff}}(t)=q_{\rm{eff}}(t)-q_0(t)$ is shown in Fig.\ \ref{fig1} by the red line. We observe that this simply estimated correction to $q_0(t)$ considerably underestimates the fully dynamical time evolution.

\begin{figure}[h]
\centering
        \includegraphics[width=0.3\textwidth, angle=-90]{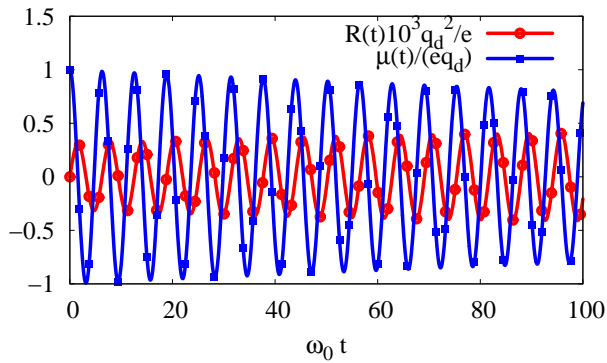}
        \caption{\label{fig3}Time dependent reaction field $R(t)$ for the time-dependent dipole moment $\mu(t)=eq(t)$ for the parameters given in Sec.\ \ref{sec.modelparam}, $\alpha=5\Gamma_0$ and $a_1/a_0=0.3$. 
}
\end{figure}

With the expectation value $q(t)=q_0(t)+q_1(t)$ at hand, we can determine the expectation value of the reaction field, given in Eq.\ (\ref{reactionfield}), in response to the dipole moment $\mu(t)$ together with Eq.\ (\ref{6}). The result is shown in 
Fig.\ \ref{fig3}. The reaction field of the surrounding medium shows a phase shift of $\pi/2$ relative to the dipole moment $\mu(t)=e q(t)$ and therefore acts as a damping force. The phase shift reflects the rearrangement time of the caging solvent molecules to the momentary configuration of the central dipole moment. 

\subsection{Influence of shrinking rate on damping}
\label{subsecRate}
To reveal the impact of the shrinking rate $\alpha$ on the damping of the central dipole, we consider the driven dynamics with a finite $q_1(t)$. Physically, a larger value of $\alpha$ brings the solvent more rapidly closer to the center of the Onsager sphere and damping is enhanced. The two limiting cases can be obtained from Eq.\ (\ref{2b}). For $\alpha \to \infty$, the damping rate $\Gamma$ is given by $\Gamma_0$, as is clear from Eq.\ (\ref{2b}). For the opposite limit $\alpha \to 0$, we may set $e^{-\alpha t} \simeq 1$ and obtain the rate $\Gamma_\alpha$ by replacing the prefactor $\frac{1}{a_0^3} \to \frac{1}{a_0^3}\left[1-\frac{3a_1}{a_0}\right]$ in the expression for $\Gamma_0$, i.e., $\Gamma \to \Gamma_\alpha = \Gamma_0 (1-3a_1 / a_0)$. 

\begin{figure}[t] 
 \begin{center}    
   \includegraphics[width=0.3\textwidth, angle=-90]{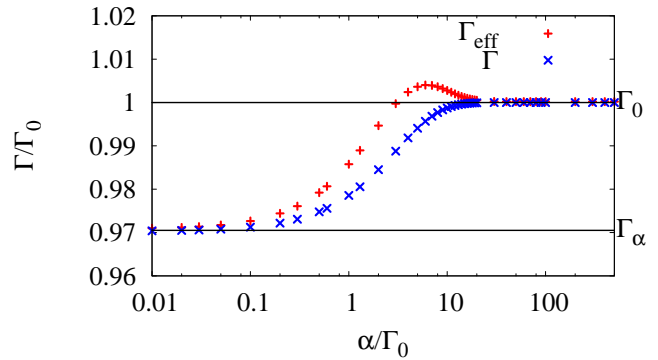}
  \caption{Damping rate $\Gamma$ (blue symbols) as a function of the shrinking rate $\alpha$ for $a_1/a_0=0.01$ and for the parameters given in Sec.\ \ref{sec.modelparam}. The limiting cases are $\Gamma \to \Gamma_0$ for $\alpha \to \infty$ and 
   $\Gamma \to \Gamma_\alpha = \Gamma_0 (1-3a_1 / a_0)$ for $\alpha \to 0$. The red symbols show the effective rate $\Gamma_{\rm eff}$ obtained from a fit to the approximated dynamics $q_{\rm eff}$. 
   }\label{fig4}
\end{center} 
\end{figure}

To obtain the crossover between the two limiting cases, we exploit the fact that the additional expectation value $q_1(t)$ has the same frequency and no phase shift to $q_0(t)$.  To determine the damping rate $\Gamma$, we fit the function $q_{\rm ft}(t)=q(t=0)e^{-\Gamma t}\cos[\omega_0t]$ to $q(t)=q_0(t) + q_1(t)$ in the regime of long times, i.e., in the time window  $[100,200]\omega_0^{-1}$, where possible fast decaying contributions are suppressed.  The result for the dependence of $\Gamma$ on the shrinking rate $\alpha$ is shown in Fig.\ \ref{fig4} for $a_1/a_0=0.01$. The full result correctly interpolates between the two limits. 

For comparison, we also show the damping rate $\Gamma_{\rm eff}$ which we obtain from fitting the decaying exponential to the adiabatic time-dependent $q_{\rm eff}(t)$ in which the static Onsager radius has been replaced by the momentary value $a(t)$. The result is shown in Fig. \ref{fig4} by the red symbols. 
We observe that the assumption of an effective dynamics leads to an overestimated damping rate as compared to the true dynamics of $q(t)$. This is consistent with the results shown in Fig.\ \ref{fig1}. Hence, we conclude that the relaxation properties of the central dipole are significantly influenced by a time-dependent radius of the Onsager sphere.

\subsection{Dipole response}
\label{subsecDipoleRes}
We now study the frequency-dependent linear response of the expectation value $\mu(t)=eq(t)$ of the dipole moment in a temporally shrinking sphere to an external harmonic electric force $eE(t)$. The response function contains information about the refraction and the absorption behavior of the test molecule. For this, we couple a time-dependent external electric field $E(t)$ to $\mu(t)$ under the assumption that the  wavelength of the external field is much larger than the sphere radius $a_0$, so that we can disregard any spatial dependence. Additionally, we assume for simplicity that the direction of $E$ is parallel $\mu$. 

To determine the response, we again perform the approximation for the prefactor $1/a(t)^3$ as above. 
\begin{figure} 
 \begin{center}    
   \includegraphics[width=0.3\textwidth, angle=-90]{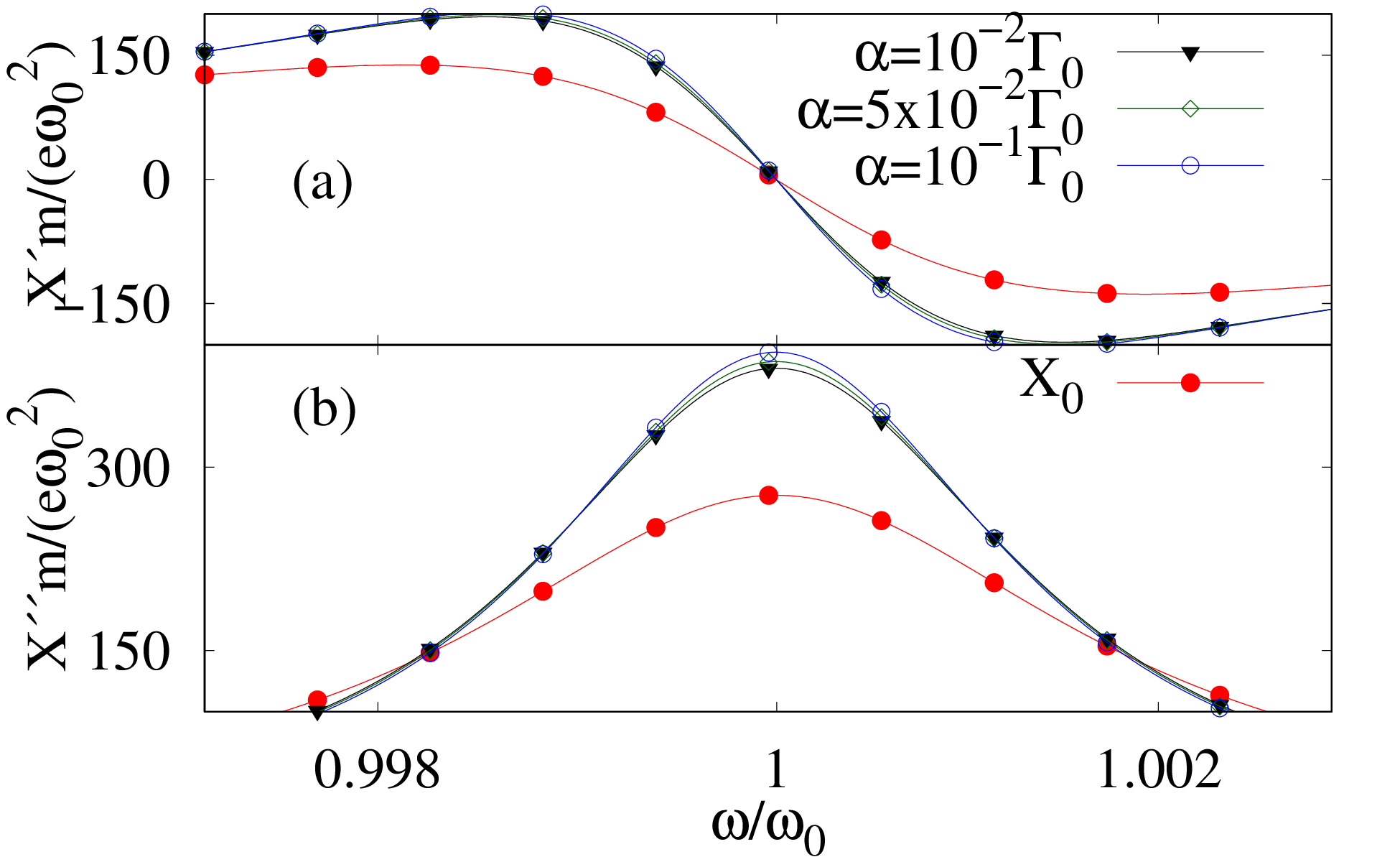}
  \caption{\label{fig5} Real (a) and imaginary (b) part of the response function $X(t=0, \omega)$ in the adiabatic limit for different shrinking rates $\alpha$ for $a_1=0.1a_0$ for the solvent parameters of water (see text).}
\end{center} 
\end{figure}
\begin{figure} 
 \begin{center}    
   \includegraphics[width=0.3\textwidth, angle=-90]{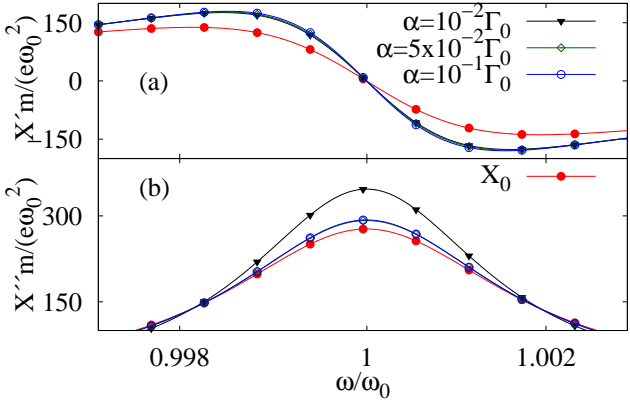}
  \caption{\label{fig6} Real (a) and imaginary (b) part of the response function $X(t=40\Gamma_0^{-1}, \omega)$ in the adiabatic limit for different shrinking rates $\alpha$ for $a_1=0.1a_0$.}
\end{center} 
\end{figure}
Next, we extend the equation of motion (\ref{1}) by the driving term and perform the Laplace transform $f(z)= i\int _{0}^{\infty} dt e^{i z t} f(t)$ to obtain $q_R(z)=q(z)+\delta q(z)$ (see Eq. \ (S10) in Suplementary Material). The first term represents the homogeneous part and is discussed in detail in the previous Sec. \ref{subsecTimevol}, while we identify the second term $\delta q(z)$ as the response term to the applied external field, i.e., 
\begin{align}
\label{38}
\delta q(z)= X_0(z)[E(z)+iW_1(z+i\alpha)E(z+i\alpha)],
\end{align}
where
\begin{align}
X_0(z)&=\frac{e}{m[\omega_0^2-z^2+i\chi_0(z)]} \, , \\ \notag
W_1(z)&=\frac{\chi_1(z)}{\omega_0^2-z^2+i\chi_0(z)}.
\end{align}
In the real-time domain and for $a_1\ll a_0$,  $\delta q(t)$ can be written in the form 
\begin{align}
\delta q(t)= i\int_0^t ds X(t,s) E(t-s),
\label{39}
\end{align}
where $s\le t$ and where 
\begin{align}
\label{40}
X(t,s)&=X_0(s)-e^{-\alpha t} \int_0^s du e^{\alpha u} X_0(u)W_1(s-u)\\ \notag &=X_0(s)-e^{-\alpha t} X_1(s).
\end{align}
\subsubsection{Adiabatic limit}
In order to obtain a frequency-dependent response function (which depends on the time $t$ as well), we again perform a Laplace transform of Eq.\ (\ref{39}), while we are only interested in the real part of $z$, which is the frequency part. According to our definition of the Laplace transform, we identify the physical response function as $X(t,\omega)\equiv-iX(t,\operatorname{Re}[z])$. The adiabatic limit holds when the shrinking rate $\alpha \ll \Gamma_0$, where $\Gamma_0$ in Eq.\ (\ref{10a}) is the decay rate of $X_0(s)$. Then, the time-dependent prefactor $e^{-\alpha t}$ can be treated adiabatically and 
the response function in the frequency domain reads
\begin{align}
X(t,\omega)=X_0(\omega)+ie^{-\alpha t} X_0(\omega-i\alpha)W_1(\omega) \, .
\label{41}
\end{align}
The response function in the adiabatic limit contains two terms. The first one reflects the response of the dipole moment to the external field for a static sphere with constant radius $a_0$. The second term includes all shrinking effects and is linear in $a_1/a_0$. Figures \ref{fig5} and \ref{fig6} show the real and imaginary part of the response function at different times $t$ after initial preparation of the dipole moment $\mu(t=0)=eq_d$.

\begin{figure} 
 \begin{center}    
   \includegraphics[width=0.3\textwidth, angle=-90]{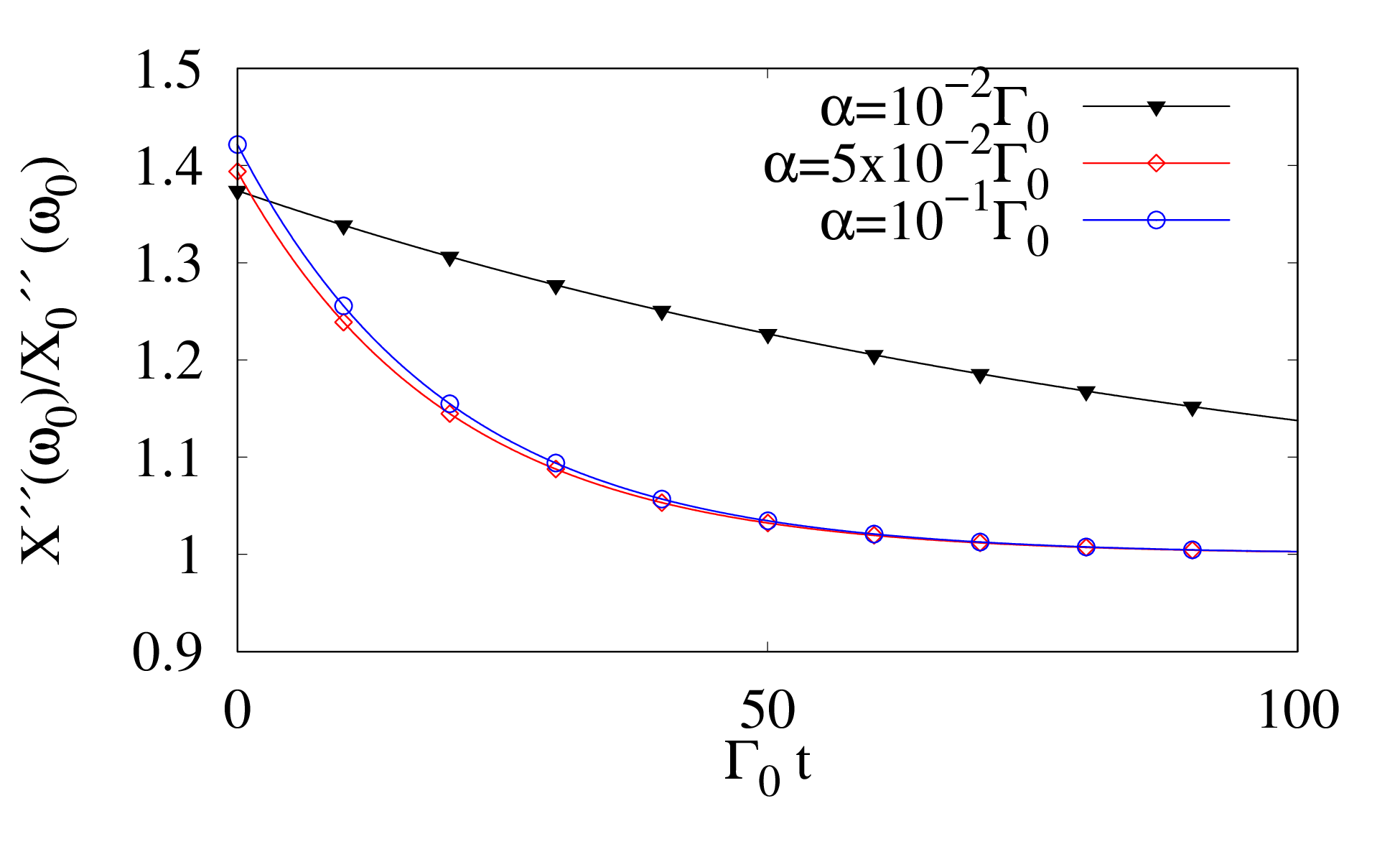}
  \caption{\label{fig7} The maximum of the imaginary part $X''(t,\omega_0)/X_0''(t,\omega_0)$ at different times for different shrinking rates $\alpha$ for $a_1=0.1 a_0$.}
\end{center} 
\end{figure}

At time $t=0$, the impact of the shrinking on the response function is most pronounced due to the second term of Eq.\ (\ref{40}). The real part of the response function is associated with the refraction of the incoming light and the imaginary part is connected to the absorption of the central dipole moment in the shrinking Onsager cavity. The latter has a maximum at $\omega_0$ and shows an enhancement of up to $40\%$ (Fig. \ref{fig5} b) for its maximum, which decays exponentially with time (Fig. \ref{fig7}).

Interestingly, a bigger shrinking rate $\alpha$ shows an initially more pronounced response which rapidly decays with time, while the response function for a smaller $\alpha$ decreases more slowly.  The physical meaning is that for a slower shrinking, the final radius $a_0$ is reached later, the action of the fluctuating environment comes later closer to the dipole, such that effectively, damping is reduced and the response is enhanced for a longer time period. In turn, the broadening of the imaginary part is narrower for longer times.

As a side remark, we may fit a Lorentzian to the imaginary part $X^{\prime \prime} (\omega)$ of Eq. (\ref{41}) and read of its full width at half maximum which is equal to $\Gamma$ for different shrinking rates $\alpha$ shown in Fig.\ \ref{fig4} for $a_1/a_0=0.01$.

\subsubsection{Beyond the adiabatic limit}

\begin{figure} 
 \begin{center}    
   \includegraphics[width=0.3\textwidth, angle=-90]{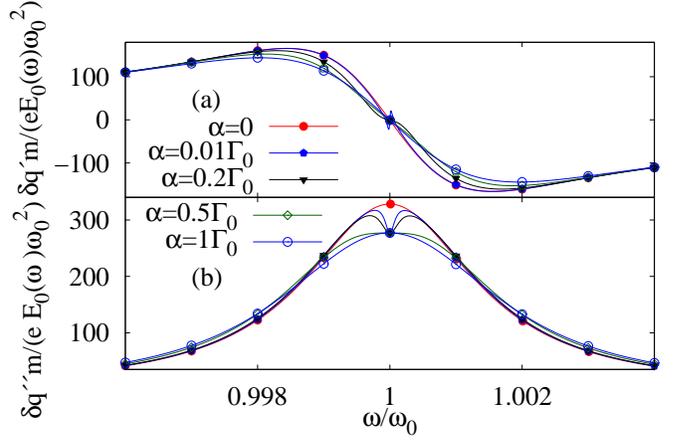}
  \caption{\label{fig8} The (a) real and (b) imaginary part of the effective response $\delta q(\omega)/E(\omega)$ for a monochromatic electric field with $\omega_{ex}=\omega_0$ for $a_1=0.1a_0$.}
\end{center}
\end{figure}

In order to get information about the dipole response beyond the adiabatic limit for larger  shrinking rates, we couple the dipole to an explicit monochromatic electric field $E(t)=E_0\cos\omega_{ex}t$. 
 We are able to calculate the response $-i\delta q(\operatorname{Re}[z])\equiv\delta q(\omega)$ according to Eq. (\ref{38}) to this specific choice of the electric field. We define the effective response function $\delta q(\omega)/ E(\omega)$ and show the results  in Fig.\ \ref{fig8}.
If the frequency of the external field is in resonance with the dipole frequency, $\omega_{ex}=\omega_0$, we find an interesting behavior. For $0\le \alpha \le 0.5 \Gamma_0$, the maximum at $\omega =\omega_0$ turns into a dip and two peaks  around the dipole eigenfrequency appear in the imaginary part of the response. 
With increasing shrinking rate, the two peaks move further apart from each other and disappear again at $\alpha\ge 0.5 \Gamma_0$ while the central linewidth becomes broader. 
Our analysis of calculating the damping rate $\Gamma$ of the dipole moment leads back to the idea of fitting a single Lorentzian to the imaginary part around the eigenfrequency $\omega_0$ which becomes broader for faster  shrinking.  

\section{Breathing Onsager Sphere}
\label{secBreath}
We now study a breathing Onsager sphere with the dynamic radius 
\begin{align}
a(t)=a_0+a_2(t)=a_0+a_2 \, \alpha \, t \, e^{1-\alpha t} \, \Theta(t) ,
\label{3}
\end{align}
which is chosen such that we obtain the maximum value $a_{\rm max}(t_{\rm max}=1/\alpha)=a_0+a_2$ at $t_{\rm max}=1/\alpha$. During this dynamics the radius first growths from $a(t=0)=a_0$ to the maximum value $a(t_{\rm max})=a_0+a_2$ and then returns to $a(t\to \infty)=a_0$. This time-dependent behavior covers excitation and de-excitation of photoexcited molecules accompanied by spatial breathing \cite{bia2016}. 

\subsection{Time evolution of the dipole moment}
\label{subsecTimeEvol2}
Next, we calculate the impact of the breathing Onsager sphere on the expectation value $\mu(t)=eq(t)$ of the central dipole moment. 
In following the same calculus as in Sec. \ref{subsecTimevol}, we again expand the prefactor $1/a(t)^3$ in a Taylor series up to first order in the time-dependent part. This is valid when the magnitude $a_2$ by which the radius changes is much smaller than the static radius $a_0$. We find 
\begin{align} \label{breathing}
\frac{1}{a(t)^3}&\simeq\frac{1}{a_0^3}\left[1-\frac{3a_2}{a_0} \alpha t e^{1-\alpha t}\right]\\ \notag
&= \frac{1}{a_0^3}\left[1-\frac{3 a_2}{a_0}\alpha \exp(1)\left(-\frac{\partial}{\partial \alpha}\right)e^{-\alpha t}\right] \, ,
\end{align}
where we can again rescale the linear the response function to $\chi_0(t)+\chi_2(t)$, since we are only interested in first order contributions of $a_2/a_0$, which is fulfilled be the prefactor of $\chi_2(t)=3a_2/a_0 \exp(1) \chi_0(t)$. Following this response function we perform a Laplace transform of Eq.\ (\ref{1}). It is helpful to define the Laplace transform of the susceptibility for the breathing according to 
\begin{equation}
\chi_2 (z)  = - \frac{3a_2}{a_0}\exp(1)\frac{e^2}{ma_0^3} \frac{\chi_D}{(\omega_D-z)^2} \, .
\label{chi2}
\end{equation} 

To obtain an analytical solution of $q(t)=q_0(t)+q_2(t)$, we need to determine the inverse Laplace transform of the dynamic part
\begin{align}
q_2(z) &=-\alpha \frac{\partial}{\partial \alpha} \frac{i\chi_2(z+i\alpha)q_0(z+i\alpha)}{\omega_0^2-z^2+i\chi_0(z)} \, , 
\end{align}
see the Supplementary Material for details. We obtain the solution 
\begin{equation}
q_2(t)=Q_2\sum_{j=1}^4 q_2^{(j)}(t),
\end{equation}
with  
\begin{equation}
Q_2 = \frac{3a_1\alpha\exp(1)}{a_0} \frac{e^2}{ma_0^3} \chi_D q(t=0) \, , 
\end{equation}
and the explicit form of the $q_2^{(j)}(t)$ as given by Eq.\ (S37) in the Supplementary Material. 

\begin{figure}[t] 
 \begin{center}    
   \includegraphics[width=0.3\textwidth, angle=-90]{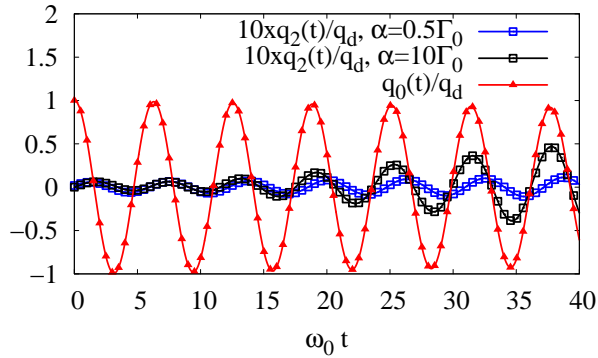}
  \caption{\label{fig9} Dynamics of the dipole moment according to $q(t)=q_0(t) + q_2(t)$. The driven part $q_2(t)$ is shown for different breathing rates $\alpha=0.5\Gamma_0$ (blue, scaled by a factor of 10) and $\alpha=10\Gamma_0$ (black, scaled by a factor of 10) for $a_2/a_0=0.3$ and for the parameters as given in Sec.\ \ref{sec.modelparam}. The undriven part $q_0(t)$ is shown in red. } 
\end{center} 
\end{figure}

The time-evolution of the dipole moment according to $q(t)=q_0(t) + q_2(t)$ is shown in Fig.\ \ref{fig9}. The driven part $q_2(t)$ is shown for different breathing rates  and compared to the undriven part $q_0(t)$. For the breathing sphere, a relative phase shift between $q_0(t)$ and $q_2(t)$ can be observed. For smaller breathing rates $\alpha/\Gamma_0<1$, the phase shift persists much longer than for the larger breathing rates.

\begin{figure}[t] 
 \begin{center}    
   \includegraphics[width=0.3\textwidth, angle=-90]{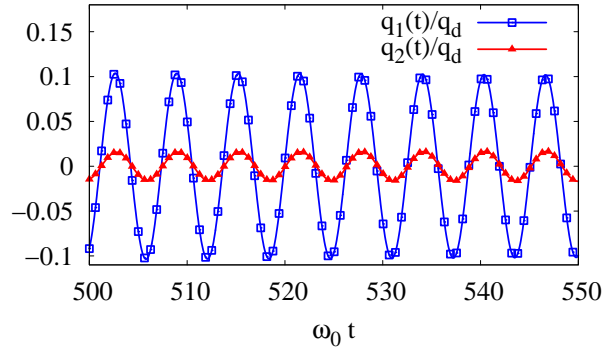}
  \caption{\label{fig10}  Dynamics  of the driven parts $q_1(t)$ (shrinking sphere, blue) and $q_2(t)$ (breathing sphere, red) at intermediate times for $a_1=a_2= 0.1 a_0$, $\alpha=0.1 \Gamma_0$, and for the parameters as given in Sec.\ \ref{sec.modelparam}. } 
\end{center} 
\end{figure}

Next, we compare the two cases of a time-dependent shrinking  $a_1(t)$ with the time-dependent breathing $a_2(t)$ Onsager radius regarding their impact on the time-evolution of $q(t)$ at longer (but still non-asymptotic) times. The comparison is shown in 
Fig.\ \ref{fig10} for the same value of the time constant $\alpha=0.1 \Gamma_0$ and for $a_{1/2}=0.1 a_0$. At long times, both cases show in-phase oscillations with $q_0(t)$. In general, the shrinking results in a more pronounced impact on the dipole oscillations than the breathing. 

\subsection{Influence of the breathing rate on damping}
\begin{figure}[t] 
 \begin{center}    
   \includegraphics[width=0.3\textwidth, angle=-90]{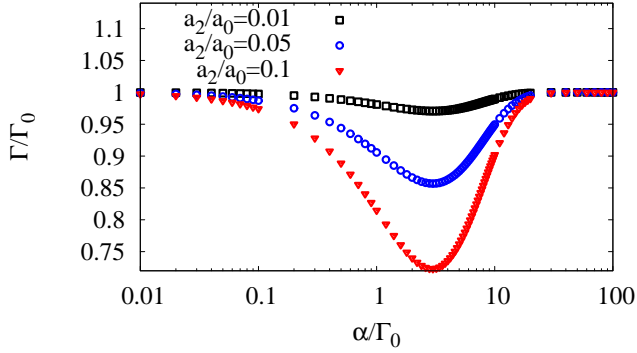}
  \caption{Dependence of the asymptotic damping rate $\Gamma$ on the breathing rate $\alpha$ 
  for different breathing amplitudes for the parameters given in Sec.\ \ref{sec.modelparam}.\label{fig11}} 
  \end{center}
\end{figure}

The parameter $\alpha$ determines the scale of the time-dependent breathing behavior of the Onsager radius $a(t)=a_0+a_2(t)$. In turn, the time scale of the damping of the dipole is given by $\Gamma_0$ and the interplay of both can affect the overall damping rate $\Gamma$ of the breathing system. In order to determine the overall damping rate $\Gamma$, we again fit the function $q_{\rm{ft}}(t)=q(t=0)e^{-\Gamma t}\cos \omega_0 t$ to $q(t)$ but now regarding the additional term which respects the breathing behavior of the Onsager radius $a(t)=a_0+a_2(t)$. The initial phase shift is neglected by restricting the fitting to the time window $\omega_0 t > 300$. The result is shown in Fig.\ \ref{fig11}. 

For small rates $\alpha \ll \Gamma_0$, the maximal extension of the sphere is reached only at long times $\sim 1/\alpha \to \infty$, so that the radius largely persists at $a_0$. Moreover, $\alpha$ enters as a prefactor in $a_2(t)$, so that the additional term of the dynamics $q_2(t)$ is negligible, and, thus, $\Gamma \to \Gamma_0$. For large breathing rates $\alpha \gg \Gamma_0$, the breathing occurs very rapidly, so that the dipole does not have enough time to respond and largely sees again an Onsager sphere with radius  $a_0$, so that $\Gamma \to \Gamma_0$. In between, when $\alpha \approx \Gamma_0$, the environmental solvent molecules are mostly pushed away from the central dipole while it relaxes, so that its damping is reduced. This effect is maximal when $\alpha = \Gamma_0$, pointing to an interesting resonant reduction of the damping. As can be seen for our choice of realistic model parameters, the reduction of the damping can amount to almost $30 \%$ in the resonance region, an effect that should be detectable by spectroscopic means. 

\subsection{Dipole response}
We next study the linear response function of the dipole moment $\mu(t)$ in the breathing sphere to an external driven electric  field $E(t)$, analogously to Sec.\ \ref{secShrink}. As before, the wavelength of the external field is assumed to be much larger than the static sphere radius $a_0$, so that we can disregard the spatial dependence of $E(t)$. Additionally, the external electric field $E(t)$ is assumed to be parallel to the dipole moment $\mu(t)$. 

The equation of motion in Laplace space again takes the form $q_R(z)=q(z)+\delta q(z)$ , where $q(z)$ is studied in detail in Sec. \ref{subsecTimeEvol2} and the term $\delta q(z)$ arises from the external field $eE(z)$. $\delta (z)$ is given by Eq. (\ref{38}), where $W_1(z)$ now takes the form 
\begin{align}
W_1(z)=-\alpha\frac{\partial}{\partial \alpha}\frac{\chi_2(z)}{\omega_0^2-z^2+i\chi_0(z)}.
\end{align}
In the real-time domain, 
\begin{align}
\label{42}
\delta q(t)=i\int_0^t ds X(t,s)E(t-s)\, ,
\end{align}
where
\begin{align}
\label{43}
X(t,s)=X_0(s)+i\alpha\frac{\partial}{\partial \alpha}e^{-\alpha t} \int_0^s du e^{\alpha u} X_0(u) W_1(s-u).
\end{align}
\subsubsection{Adiabatic limit}
When performing the Laplace transform of $X(t,s)$, one again assumes $\alpha\ll \Gamma_0$, the decay rate of $X_0(s)$. Thus, the time-dependent prefactor can be treated adiabatically. The Laplace transform of the response function for the breathing sphere then becomes 
\begin{align}
X(t,\omega)&=X_0(\omega)+i\alpha t e^{-\alpha t} X_0(w-i\alpha)W_1(\omega)\\ \notag 
&-i\alpha e^{-\alpha t}\frac{\partial}{\partial \alpha}X_0(\omega-i\alpha)W_1(\omega)\, ,
\end{align}
where the real part of the Laplace variable $\operatorname{Re}[z]\equiv \omega$ reflects the frequency. According to our definition of the Laplace transform, the physical response is identified as $X(\omega)\equiv-iX(\operatorname{Re}[z])$.
\begin{figure}[t] 
 \begin{center}    
   \includegraphics[width=0.25\textwidth, angle=-90]{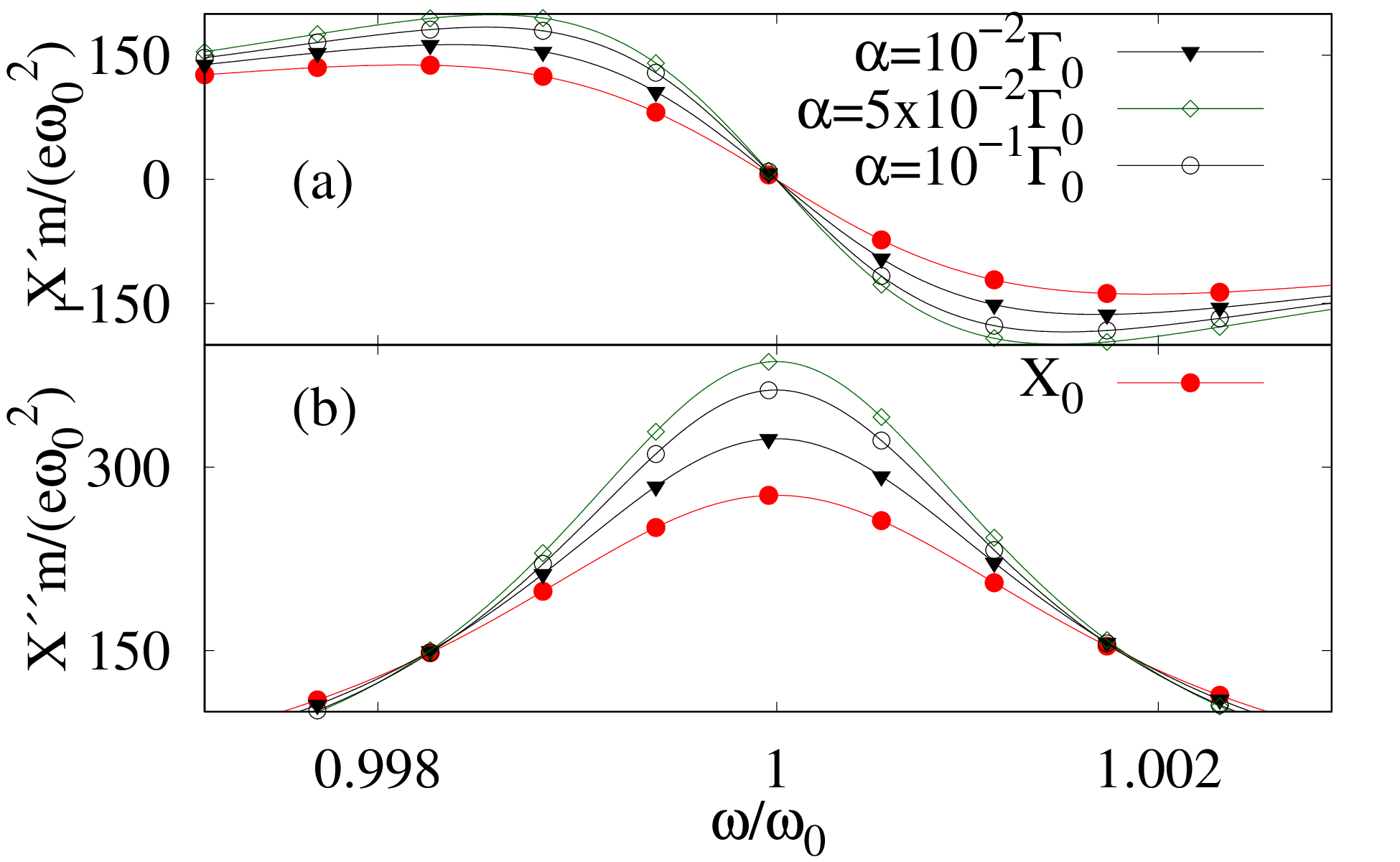}
  \caption{\label{fig12}Real (a) and imaginary (b) part of the response function $X(t=20\Gamma_0^{-1},\omega)$ for different breathing rates $\alpha$ for $a_2=0.1 a_0$. Note that $\Gamma_0=22.2\times 10^{-4}\omega_0$ for the parameters given in Sec.\ \ref{sec.modelparam}.}
\end{center} 
\end{figure}
At initial time $t=0$ the response of the dipole moment in a breathing sphere does not differ from the dipole moment in the static sphere. We show the results for a later time $t=20\Gamma_0^{-1}$ in Fig. \ref{fig12}.  One observes again a more pronounced response function which finally drops to the form of a static sphere (Fig.\  \ref{fig13}, e.g., the imaginary part). Especially the imaginary part which reflects the absorption of the dipole moment is temporally enhanced by up to $40\%$ as compared to the static case. A smaller breathing rate $\alpha$ shows a smaller impact on the response. Additionally the broadening of the line shape of the absorptive part is reduced at different times for different rates $\alpha$. A temporal smaller broadening is connected to a momentary reduced damping of the central dipole moment. The maximum radius $a(t=1/\alpha)=a_0+a_2$ of the breathing sphere is reached at the time $t=1/\alpha$ where the fluctuating environment is spatially maximal far away from the dipole moment, while its damping is the most reduced.
Thus, the time-dependent response function carries clear signatures of the breathing of the Onsager sphere.

\subsubsection{Beyond the adiabatic limit}

The effective response $\delta q(\omega)/E(\omega)$  to an electric field $E(t)=E_0\cos\omega_{ex} t$ leads also to a split in two peaks shifted around the dipole eigenfrequency $\omega_0$, while the central peak remains the most pronounced. This splitting first increases with enhanced breathing rate until it disappears at $\alpha \ge 0.4\Gamma_0$. At $\alpha \approx \Gamma_0$ we observe the narrowest line width which eventually becomes broader with increasing breathing rate, see Fig. \ref{fig14}. This can be again recovered by recording the full width at  half maxmimum of a single Lorentzian fit to the imaginary part, see discussion in Sec.\ \ref{subsecDipoleRes}. These results are recorded in Fig.\ \ref{fig11} as fitted damping rates to the real-time dipole dynamics $q(t)$.
\begin{figure}[t] 
 \begin{center}    
   \includegraphics[width=0.25\textwidth, angle=-90]{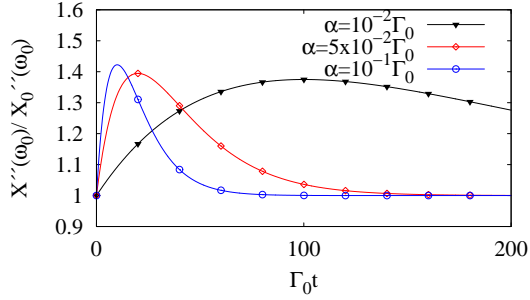}
  \caption{\label{fig13}Time-dependent maximum of $X^{\prime \prime} (t,\omega_0)$ of the response function for $a_2=0.1 a_0$ for different breathing rates.}
\end{center} 
\end{figure}

\begin{figure} 
 \begin{center}    
   \includegraphics[width=0.3\textwidth, angle=-90]{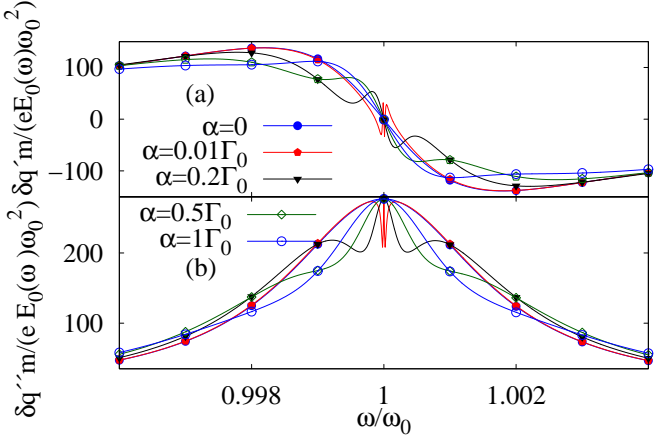}
  \caption{\label{fig14} The (a) real and (b) imaginary part of the effective response $\delta q(\omega)/E(\omega)$ for a monochromatic electric field at eigenfrequency $\omega_{ex}=\omega_0$ for $a_1=0.1a_0$.}
\end{center}
\end{figure}

\section{Conclusions} \label{secConcl}
We establish the time-dependent Onsager model in order to portray a solvated molecule which changes its spatial extensions in a time-dependent way. As an important consequence, the interaction of the molecular dipole moment with the dissipative solvent becomes time-dependent, such that the effects of nonequilibrium quantum solvation can be investigated within a model-based approach. For this, we have assumed an external time-dependent forcing such that the radius of the Onsager sphere varies over time. We consider the cases of a shrinking and a breathing Onsager sphere. The solvent provides a polar environment which forms a delayed reaction electric field which couples to the molecular dipole moment and hence damps its oscillating dynamics. We have determined the time-dependent response of the dipole in terms of its relaxational properties to the time-dependent Onsager sphere. For this, we have considered the relative change in the radius of the sphere over time as a perturbation, such that the global response of the dipole can be calculated analytically for all values of the time constants of the changes, when the dipole oscillations are described in terms of a damped harmonic oscillator. 

For the case of a shrinking sphere, we find that the qualitative form of the response is close to the that of the static sphere, but is strongly time-dependent. The magnitude of the absorptive part is temporally enhanced up to $40\%$. An effective response to an explicit external electric field induces a splitting of the central peak in the absorptive part which increases with enhanced shrinking. As this splitting is small, the oscillations of the dipole in the shrinking sphere have nearly the same phase as those of the static sphere. The damping rate can be easily understood in the two limits of fast and slow shrinking and our results nicely interpolate between these two limits. Our results clearly show that the nonequilibrium quantum solvation cannot be described by an adiabatic approach which treats the momentary radius of the shrinking sphere as a parameter. In general, a shrinking sphere shows a stronger damping than the corresponding static sphere.

For the case of a breathing sphere, the real and the imaginary parts of the susceptibility show again a strong time dependence. A breathing Onsager sphere is accompanied by a globally reduced damping at intermediate shrinking rates. There, the damping rate shows a significant minimum for values of shrinking rates comparable to the static relaxation relaxation. This cross-over behavior nicely interpolates between the two limits of fast and slow breathing which both coincide with the relaxation behavior of a static sphere. 

The resonantly reduced damping which occurs for a breathing Onsager sphere illustrates the nontrivial role of a nonequilibrium environment for the relaxation properties of a damped system. It can only occur under non-equilibrium conditions and shows that a time-dependent dissipative environment away from thermal equilibrium can yield to less detrimental implications than resulting from a thermal reservoir. The time-dependent Onsager model thus contributes to understanding the impact of dynamically and spatially varying molecular properties of dissolved molecules on the relaxation of its electronic degree of freedom. The task remains to reveal the consequences of a nonequilibrium environment for the dynamics of the dipole moment by molecular electronic spectroscopy. In particular, nonlinear electronic spectroscopy could provide insight into the interplay of the dipolar dynamics and its nonequilibrium environment. 

\section*{Supplementary Material}
See Supplementary Material for a details on the equation of motion and the explicit time-dependent forms of the time-dependent dipole oscillations. 

\section*{Acknowledgments}
We are grateful to Christian Bressler for helpful discussions. This work is supported by the DFG Sonderforschungsbereich 925 (Project A4).

\newpage



\begin{thebibliography}{99}
\bibitem{Review}J. Tomasi, B. Mennucci, and R. Cammi, Chem. Rev. {\bf 105}, 2999 (2005).

\bibitem{MayBook}V. May and O. K\"uhn, \textit{Charge and Energy Transfer Dynamics in Molecular Systems}, 3rd ed.\ (Wiley-VCH, Weinheim, 2011).

\bibitem{NitzanBook}A. Nitzan, \textit{Chemical Dynamics in Condensed Phases} (Oxford University Press, Oxford, 2006).


\bibitem{bot1973}C. J. F. B\"ottcher, \textit{Theory of electric polarization} (Elsevier, Amsterdam, 1973).

\bibitem{bor1920}M. Born, Z.\ Phys.\ \textbf{1}, 55 (1920).

\bibitem{kir1935}J. G. Kirkwood, J. Chem. Phys. \textbf{3}, 300 (1935).

\bibitem{ons1936}L. Onsager, J. Am. Chem. Soc. \textbf{58}, 1486 (1936).

\bibitem{wei2012}U. Weiss, \textit{Quantum Dissipative Systems} (World Scientific Publishing, Singapore, 2012).

\bibitem{leg1987}A. J. Leggett, S. Chakravarty, A. T. Dorsey, M. P. A. Fisher, A. Garg, and W. Zwerger, Rev. Mod. Phys. \textbf{67}, 1 (1987).

\bibitem{mck2005}J. Gilmore and R. H. McKenzie, J.Phys.: Condens. Matter \textbf{17}, 1735 (2005).

\bibitem{nal2014} P. Nalbach, A.J.A. Achner, M. Frey,  M. Grosser, C. Bressler, and M. Thorwart, J. Chem. Phys. \textbf{141}, 044304 (2014).

\bibitem{bre2014}C. Bressler, W. Gawelda, A. Galler, M. M. Nielsen, V. Sundstr\"om, G. Doumy, A. M. March, S. H. Southworth, L. Young, and G. Vank\'{o}, Faraday Discuss. \textbf{171}, 169 (2014).


\bibitem{pen2012}T. J. Penfold, B. F. E. Curchod, I. Tavernelli, R. Abela, U. Rothlisberger, and M. Chergui, Phys. Chem. Chem. Phys.\textbf{14}, 9444 (2012).

\bibitem{dak2010}L. M. L. Daku and A. Hauser, J. Phys. Chem. Lett. \textbf{12}, 1830 (2010).

\bibitem{pen2013}T. J. Penfold, C. J. Milne, I. Tavernelli, and M. Chergui, Pure Appl. Chem. \textbf{85}, 53 (2013).

\bibitem{Corni1}S. Corni, S. Pipolo, and R. Cammi, J. Phys. Chem. A {\bf 119}, 5405 (2014). 

\bibitem{Corni2}S. Pipolo, S. Corni, and R. Cammi, J. Chem. Phys. {\bf 146}, 064116 (2017).

\bibitem{bia2016}E. Biasin, T. B. van Driel,  K. S. Kj\ae{}r, A. 0. Dohn, M. Christensen, T. Harlang, P. Chabera, Y. Liu, J. Uhlig, M. P\'apai, Z. N\'emeth, R. Hartsock, W. Liang, J. Zhang, R. Alonso-Mori, M. Chollet, J. M. Glownia, S. Nelson, D. Sokaras, T. D. Assefa, A. Britz, A. Galler, W. Gawelda, C. Bressler, K. J. Gaffney, H. T. Lemke, K. B. M\o{}ller, M. M. Nielsen, V. Sundstr\"om, G. Vank\'o, K. W\"arnmark, S. E. Canton, K. Haldrup, Phys. Rev. Lett. \textbf{117}, 013002 (2016).

\bibitem{del2014}M. Delor, P. A. Scattergood, I. V. Sazanovich, A. W. Parker, G. M. Greetham, A. J. H. M. Meijer, M. Towrie, and J. A. Weinstein, Science \textbf{346}, 1492 (2014).

\bibitem{lin2009}Z. Lin, C. M. Lawrence, D. Xiao, V. V. Kireev, S. S. Skourtis, J. L. Sessler, D. N. Beratan, and I. V. Rubtsov, J. Am. Chem. Soc. \textbf{131}, 18060 (2009).

\bibitem{lyn2012}M. S. Lynch, K. M. Slenkamp, and M. Khalil, J. Chem. Phys. \textbf{136}, 241101 (2012).

\bibitem{drivenbath1}H. Grabert, P. Nalbach, J. Reichert, and M. Thorwart, 
J. Phys. Chem. Lett. {\bf 7}, 2015 (2016).

\bibitem{drivenbath2}J. Reichert, P. Nalbach, and M. Thorwart, 
Phys. Rev. A {\bf 94}, 032127 (2016). 

\bibitem{jac1999}J. D. Jackson, \textit{Classical Electrodynamics} (Wiley, New York, 1999).

\bibitem{weg2004}M. Wegner, \textit{Extreme Nonlinear Optics} (Springer, Berlin, 2004).

\end{thebibliography}
\end{document}



\title{Supplementary Material to \\[3mm] 
Nonequilibrium quantum solvation with a time-dependent Onsager cavity}


\author{H. Kirchberg}
\affiliation{I. Institut f\"ur Theoretische Physik, Universit\"at Hamburg, Jungiusstra\ss{}e 9, 20355 Hamburg, Germany}
\author{P. Nalbach}
\affiliation{Westf\"alische Hochschule, M\"unsterstr. 265, 46397 Bocholt, Germany}
\author{M. Thorwart}
\affiliation{I. Institut f\"ur Theoretische Physik, Universit\"at Hamburg, Jungiusstra\ss{}e 9, 20355 Hamburg, Germany}

\date{\today}

\maketitle

\section{Overview}
In Sec.\ \ref{EQM}, we rewrite the equation of motion in the usual form in which a Stokes damping force appears which is proportional to the velocity of the particle. In Sec.\ \ref{LBT}, we provide the details of the inverse Laplace transform to obtain the time-dependent response function and the solution of the equation of motion for the case of the shrinking Onsager sphere in the time domain. The corresponding solution for the case of the breathing Onsager sphere is given in Sec.\ \ref{breathing}.

\section{Equation of motion in time domain}
\label{EQM}
The expectation value $\langle \mu(t) \rangle =e \langle q(t) \rangle$ of the dipole moment embedded in the center of the Onsager sphere follows from the solution of the equation of motion in which the expectation value of the reaction field $\langle R(t) \rangle=\frac{e^2}{a(t)^3}\int_0^t dt' \chi(t-t')\langle q(t') \rangle$ acts as an  external force $e \langle R(t) \rangle$ on the dipole moment. For better readability, we set $\langle q(t) \rangle \equiv q(t)$ and $\langle R(t) \rangle \equiv R(t)$. The equation of motion then reads
\begin{align}
m\ddot{q}(t)+m\omega_0^2q(t)-\frac{e^2}{a(t)^3}\int_0^t dt' \chi(t-t')q(t')=0,
\label{31}
\end{align}
where the third term is the back action of the dielectric medium on the dipole.
To obtain the usual form of the equation of motion with a damping force proportional to the velocity we express the response function $\chi(t)$  in terms of the damping kernel 
\begin{align}
\gamma(t)=\int_t^\infty ds \chi(s)=\frac{\chi_D}{\omega_D}e^{-\omega_D t} \, ,
\end{align}
where we have used the Debye form of Eq.\ (6) of the main text for the susceptibility. With this, we can now write the third term of Eq. (\ref{31}) in terms of the damping kernel in the form 
\begin{equation}
m\ddot{q}(t)+m\omega_0^2q(t)+\frac{e^2}{a(t)^3}\int_0^t dt' \dot{\gamma}(t-t')q(t')=0 \, .
\end{equation}
After an integration by parts and obeying the relation $\dot{\gamma}(t-t')=\partial\gamma(t-t')/\partial(t-t')=-\partial\gamma(t-t')/\partial t'$, we find 
\begin{align}
m\ddot{q}(t)+m\omega_0^2q(t)-\frac{e^2}{a(t)^3}\left[\gamma(0)q(t)-\gamma(t)q(0)-\int_0^t dt'\gamma(t-t')\dot{q}(t')\right]=0\, .
\end{align}
or, equivalently, 
\begin{align}
\label{32}
m\ddot{q}(t)+m\left[\omega_0^2-\frac{e^2}{ma(t)^3}\gamma(0)\right]q(t)+\frac{e^2}{a(t)^3}\int_0^t dt'\gamma(t-t')\dot{q}(t')=-\frac{e^2}{a(t)^3}\gamma(t)q(0).
\end{align}
The second term describes a renormalized potential where the renormalization occurs due to the coupling to the dielectric medium, where the effective frequency follows from  $\omega_{\rm eff}^2=\omega_0^2-[\Delta \omega(t)]^2$, with $[\Delta \omega (t)]^2=e^2 \gamma(0)/(ma(t)^3)$. The biggest renormalization occurs for the smallest magnitude of the sphere radius $a_0$ such that we set
\begin{equation}
\Delta \omega \equiv\Delta \omega(t\to \infty) = \sqrt{\frac{\Phi}{\omega_D}} \, , 
\end{equation}
where we have defined
\begin{equation}
\Phi = \frac{e^2}{ma_0^3} \chi_D\, .
\end{equation}
If $\omega_0^2<(\Delta\omega)^2$, the effective potential is has a negative curvature and describes an unstable situation of the dipole moment in the Onsager sphere. So, 
since we are only interested in the damping of the dynamics of the dipole induced by the dielectric medium and not in an unstable renormalization of the potential, we need to ensure that 
$\Delta \omega < \omega_0$ which is fulfilled for our choice of parameters in the main text.  

The third term of Eq.\ (\ref{32}) describes the damping force acting on the dipole in the Onsager sphere due to the response of the environment to the dipolar dynamics. This damping kernel is related to the spectral density $J(\omega)=m\omega\int_0^\infty dt \gamma(t) \cos(\omega t)$ in case of a static Onsager radius $a(t)=a_0$ of an equilibrium environment described as an infinite set of harmonic ocillators \cite{wei2012}. The inhomogeneous term on the right-hand side of Eq.\ (\ref{32}) is an initial slip resulting from the present assumption of factorizing initial conditions for the system-bath dynamics \cite{wei2012}. 

\section{Dynamics of the shrinking Onsager sphere}
\label{LBT}
The equation of motion (9) in the main text is a linear integro-differential equation, which cannot be solved in an analytical manner for a general form of $a(t)$. To employ the Laplace transform for its solution, we need to use an approximation for the pefactor $1/a(t)^3=1/(a_0+a_1(t))^3$. Here, we consider the case of a small variation of the Onsager sphere by an overall magnitude $a_1\ll a_0$, by which the radius changes. In fact, the molecule is typically much larger than the change of its size induced by photoexcitation. Then, we can Taylor expand the prefact to first order in $a_1$ and find 
\begin{align}
\frac{1}{a(t)^3}=\frac{1}{(a_0+a_1e^{-\alpha t})^3}\simeq \frac{1}{a_0^3}\bigg[1-3\frac{a_1}{a_0}e^{-\alpha t}\bigg].
\end{align} 
In this limit and by introducing the rescaled susceptibilities of Eq.\ (12) in the main text, we obtain the equation of motion in Laplace space as 
\begin{align}
\label{38}
q(z)= q_0(z)+q_1(z) &= \frac{zq(t=0)}{\omega_0^2-z^2+i\chi_0(z)} +\frac{i\chi_1(z+i\alpha)q(z+i\alpha)}{\omega_0^2-z^2+i\chi_0(z)}\\ \notag &\simeq \frac{z q(t=0)}{\omega_0^2-z^2+i\chi_0(z)} +\frac{i\chi_1(z+i\alpha)q_0(z+i\alpha)}{\omega_0^2-z^2+i\chi_0(z)},
\end{align}
where $q_0(z)$ includes the contribution form the static smaller sphere with the final radius $a_0$, while all contributions from the time dependent radius are contained in $q_1(z)$. Here, we make sure that the second term $q_1(z)$ only contains contributions up to first order in $a_1/a_0$, which is realized by $\chi_1(z+i\alpha)$.  Therefore, we need to replace $q(z+i \alpha)$ by $q_0(z+i \alpha)$. 

We are now able to transform the terms $q_0(z)$ and $q_1(z)$ via an inverse Laplace transform into time domain. The eigenfrequency for $q_0(z)$ is chosen to describe a stable harmonic potential, i.e., $\omega_0>\Delta\omega$. 
\subsection{Response function}
When we add an external force $eE(t)$ on the rhs of Eq.\ (12) and then perform a Laplace transform, we find 
\begin{align}
q_R(z)&=q(z)+\delta q(z) \\ \notag &= \frac{z q(t=0)}{\omega_0^2-z^2+i\chi_0(z)} +\frac{i\chi_1(z+i\alpha)q_0(z+i\alpha)}{\omega_0^2-z^2+i\chi_0(z)} \\ \notag &+ \frac{e}{m}\frac{1}{\omega_0^2-z^2+i\chi_0(z)}\bigg[E(z) +\frac{i\chi_1(z+i\alpha)E(z+i\alpha)}{\omega_0^2-(z+i\alpha)^2+i\chi_0(z+i\alpha)}\bigg ],
\end{align} 
while we again replace $q(z+i\alpha)$ by $q_0(z+i\alpha)$ for the same reason as above. The second term $\delta q(z)$ reflects the response of the dipole moment to the external force and reads
\begin{align}
\delta q(z)=X_0(z)[E(z)+iW_1(z+i\alpha)E(z+i\alpha)],
\end{align} 
with 
\begin{align}
X_0(z)&=\frac{e}{m[\omega_0^2-z^2+i\chi_0(z)]}, \\ 
W_1(z)&=\frac{\chi_1(z)}{\omega_0^2-z^2+i\chi_0(z)}.
\end{align}
We define the Laplace transform
\begin{align}
\label{LT}
f(z)=i\int_0^{\infty} dz e^{izt} f(t) = L[f(t)],
\end{align}
where $z \in \mathbb{C}$ and $\operatorname{Im}[z]>0$, so that the convolution rule follows as
\begin{align}
L[\int_0^t ds f(t-s)q(s)]=-if(z)q(z).
\end{align}
In the real-time domain, $\delta q(t)$ takes the form (for small but finite radius $a_1$)
\begin{align}
\delta q(t)= i \int_0^t ds X_0(t-s)[E(s)-e^{-\alpha s} \int_0^s du W_1(s-u)E(u)]\, .
\end{align}
This may be written as 
\begin{align}
\delta q(t)=i\int_0^t X(t,s)E(t-s),
\end{align}
where
\begin{align}
X(t,s)=X_0(s)-e^{-\alpha t}\int_0^s du e^{\alpha u}X_0(u)W_1(s-u)=X_0(s)-e^{-\alpha t} X_1(s).
\end{align}
The Fourier transform of $X(t,s)$ leads to the response function. An explicit form of it can only be obtained in the adiabatic limit: One assumes $\alpha\ll \Gamma_0$, while $\Gamma_0$ is the decay rate of $X_0(s)$, so that the time-dependent prefactor $e^{-\alpha t}$ can be treated adiabatically. We define the Fourier transform as
\begin{align}
f(\omega)=\int_{-\infty}^{\infty} dt e^{i\omega t} f(t) = F[f(t)].
\end{align}
As all functions $f(t)=0$, $t<0$ in our problem, one performs a Laplace transform, so that one may write $f(\omega)\equiv-if(\operatorname{Re}[z])$ in view of our definition of the Laplace transform in Eq.\ (\ref{LT}).

\subsection{Static part}
To transform $q_0(z)$ back into the time domain, one first calculates the roots of the denominator polynomial $p(z)$ according to 
\begin{align}
q_0(z)&=\frac{q(t=0)z}{\omega_0^2-z^2+i\chi_0(z)}=\frac{q(t=0)(-z^2-iz\omega_D)}{z^3+i\omega_D z^2-\omega_0^2z+i(-\omega_0^2 \omega_D+\Phi)}=\frac{q(t=0)(-z^2-iz\omega_D)}{D(z)} \label{4}.
\end{align}

The resulting equation 
\begin{align}
D(z)=z^3+i\omega_D z^2-\omega_0^2z+i(-\omega_0^2 \omega_D+\Phi)=0,
\end{align}
can now be solved in several steps. First, we substitute $z=i\lambda$ and obtain 
\begin{align}
\lambda^3+\omega_D \lambda^2+\omega_0^2\lambda+\omega_0^2 \omega_D-\Phi=0\, .
\end{align}
With the shift $\lambda=\tilde{z}-\omega_D/3$, it follows that 
\begin{align}
\tilde{z}^3+p\tilde{z}+r=0,
\end{align}
where $p=\omega_0^2-\omega_D^2/3$ and $r=2\omega_D^3/27+2\omega_0^2\omega_D/3-\Phi$.
The cubic equation can be solved by Cardano's method and one obtains
\begin{align}
\label{3}\tilde{z}_1&=u+v\, , \\
\tilde{z}_2&=u\epsilon_1+v\epsilon_2 \, , \\
\tilde{z}_3&=u\epsilon_2+v\epsilon_1,
\end{align}
where $\epsilon_1=-1/2+i\sqrt{3}/2$, $\epsilon_2=-1/2-i\sqrt{3}/2$, $u=(-r/2+\sqrt{\Delta})^{1/3}$, $v=(-r/2-\sqrt{\Delta})^{1/3}$ and $\Delta=(r/2)^2+(p/3)^3$. 

Next, we exploit the relations between the parameters which are typical for a Debye solvent, and, in this case, for water in the Debye approximation. Plugging in the parameters discussed in Sec.\ III in the main text, we find that $\Phi \approx \omega_0^2 \omega_D$ and that $p\approx \omega_0^2$. Moreover, we have that $r\approx 2\omega_0^2 \omega_D / 3 -\Phi \approx -\omega_0^2 \omega_D/3<0$. This ensures that $\Delta \simeq \omega_0^6/3^3 >0$, such that three linearly independent solutions exist, since $\omega_0 \gg \omega_D$.   

After back-substitution, the roots read
\begin{align}
\label{z1} z_1&=i\left(u+v-\frac{\omega_D}{3}\right)\equiv-i c \, , \\
\label{z2} z_2&=-\frac{i}{2}\left(u+v+\frac{2\omega_D}{3}\right)-\frac{\sqrt{3}}{2}(u-v)\equiv-i\Gamma_0-\omega \, ,\\
\label{z3}z_3&=-\frac{i}{2}\left(u+v+\frac{2\omega_D}{3}\right)+\frac{\sqrt{3}}{2}(u-v)\equiv-i\Gamma_0+\omega\, .
\end{align}
Here, we have introduced the parameters
\begin{align}
\label{c1} c&=-u-v+\frac{\omega_D}{3} \, , \\
\label{c2} \Gamma_0&=\frac{1}{2}\left(u+v+\frac{2\omega_D}{3}\right)\, ,\\
\Omega&=\frac{\sqrt{3}}{2}(u-v)\, .
\end{align}
Using again the physically motivated approximations, we have that $u\simeq \omega_0/\sqrt{3}$ and $v=-u$, such that 
\begin{align}
\label{d1} c&\simeq\frac{\omega_D}{3} \, , \\
\label{d2} \Gamma_0&\simeq\frac{\omega_D}{3}\, ,\\
\Omega&\simeq\omega_0\, .
\end{align}

%

The inverse Laplace transformation can be carried out by solving 
\begin{align}
\label{q0}
q_0(t)=\frac{1}{2\pi i}\int_{-\infty}^{\infty} dz e^{-izt}\frac{q(t=0)(-z^2-i\omega_D z)}{(z-z_1)(z-z_2)(z-z_3)}=-2\pi i \sum_j {\rm Res}(q_0, z_j)\, .
\end{align}
We note that due to $e^{-izt}$ and $t>0$, we need to close the integration contour in the lower half complex plane  in order to use the theorem of residue which yield the additional minus sign.

In time-domain $q_0(t)$ finally reads

\begin{align}
\label{30}
q_0(t)&\simeq e^{-\Gamma_0 t} q(t=0) \left[
\frac{2}{9} \frac{\omega_D^2}{\omega_0^2} + \left(\frac{2}{9} \frac{\omega_D^2}{\omega_0^2}+1\right) \cos \omega_0 t - \frac{\omega_D}{3\omega_0} \sin \omega_0 t
\right] \\
&\simeq e^{-\Gamma_0 t} q(t=0)\cos \omega_0 t\, .
\end{align}

\subsection{Dynamic part}

The dynamic part  $q_1(z)$ can be expressed as
\begin{align}
q_1(z)&= \frac{i\chi_1(z+i\alpha)q_0(z+i\alpha)}{\omega_0^2-z^2+i\chi_0(z)} \\ \notag
&=i Q_1 \frac{1}{z-z_0} \frac{z+i\omega_D}{(z-z_1)(z-z_2)(z-z_3)}
\frac{z^2+i\omega_D z - \alpha^2 - \omega_D \alpha +2i\alpha z}{(z-\bar{z}_1)(z-\bar{z}_2)(z-\bar{z}_3)} \, ,
\end{align}
where $\bar{z}_i = z_i -i \alpha$ are the shifted roots $z_i$ of Eqs.\ (\ref{z1})-(\ref{z3}), 
\begin{equation} \label{z0}
z_0 = -i (\alpha + \omega_D) \, ,
\end{equation}
and 
\begin{equation}
Q_1 = \frac{3a_1}{a_0} \frac{e^2}{ma_0^3} \chi_D q(t=0) \, .
\end{equation}

After performing the inverse Laplace transform back to the time domain by using the theorem of residues, we obtain three terms such that 
\begin{equation}
q_1(t) = Q_1 \sum_{j=1}^3 q_1^{(j)}(t) \, ,
\end{equation}
with
\begin{align}
q_1^{(1)}(t) &=  e^{-\omega_D t/3} \, \frac{1}{3}\, \frac{\omega_D}{\omega_0^2}  \, 
\frac{1}{\alpha (\alpha^2 + \omega_0^2)}
\left\{
\frac{2}{3}\omega_D +\alpha - \frac{\alpha(5\omega_D-9\alpha)}{2\omega_D + 3 \alpha}
-e^{-\alpha t}\left(\frac{2}{3}\omega_D -\alpha\right) 
\right\}\, , \notag \\
q_1^{(2)}(t) &= e^{-\omega_D t /3} \frac{1}{9\omega_0^2 \alpha} {\rm Re} \left\{
e^{i\omega_0 t} \frac{3\omega_0-i 2 \omega_D}{3\omega_0-i (2\omega_D+3\alpha)}
\frac{2\omega_D^2+9\omega_0^2-3\omega_D\alpha-9\alpha^2-i(3\omega_D\omega_0+18\alpha\omega_0)}{(i\alpha-\omega_0)(i\alpha-2\omega_0)}
\right\}\, , \notag \\
q_1^{(3)}(t) &= e^{-(\omega_D/3+\alpha) t }\frac{1}{9\omega_0^2 \alpha}    {\rm Re} \left\{
e^{i\omega_0 t } \frac{3\omega_0-i (2\omega_D-3\alpha)}{3\omega_0-i 2\omega_D}
\frac{2\omega_D^2+9\omega_0^2-i 3\omega_D\omega_0}{ (i\alpha+\omega_0)(i\alpha+2\omega_0)}
\right\} \, .
\end{align}

\section{Dynamics of the breathing Onsager Sphere}
\label{breathing}
To obtain an analytical solution of $q(t)=q_0(t)+q_2(t)$ in the case of a breathing Onsager sphere, we need to determine the inverse Laplace transform of the dynamic part
\begin{align}
q_2(z) &=-\alpha \frac{\partial}{\partial \alpha} \frac{i\chi_2(z+i\alpha)q_0(z+i\alpha)}{\omega_0^2-z^2+i\chi_0(z)} \\ \notag
&=-i Q_2 \frac{\partial}{\partial \alpha}\bigg[\frac{1}{z-z_0} \frac{z+i\omega_D}{(z-z_1)(z-z_2)(z-z_3)}
\frac{z^2+i\omega_D z - \alpha^2 - \omega_D \alpha +2i\alpha z}{(z-\bar{z}_1)(z-\bar{z}_2)(z-\bar{z}_3)}\bigg] \, ,
\end{align}
where $\bar{z}_i = z_i -i \alpha = \bar{z}_i(\alpha)$ are the shifted roots $z_i$ of Eqs.\ (\ref{z1})-(\ref{z3}) and explicitly depend on $\alpha$. Moreover, $z_0$ is given by Eq.\ (\ref{z0}) and also depends on $\alpha$. Furthermore,  
\begin{equation}
Q_2 = \frac{3a_1\alpha\exp(1)}{a_0} \frac{e^2}{ma_0^3} \chi_D q(t=0) \, .
\end{equation}
Performing the inverse Laplace transform by exploiting the theorem of residues, one obtains the solution $q(t)=q_0(t)+q_2(t)$ in the time domain, where $q_0(t)$ is known from Eq. (\ref{q0}). The solution $q_2(t)$ reads 
\begin{equation}
q_2(t)=Q_2\sum_{j=1}^4 q_2^{(j)}(t),
\end{equation}
with
\begin{align}
q_2^{(1)}(t) &=  e^{-\omega_D t/3} \, \frac{2}{3} \frac{\omega_D}{\omega_0^2}  \, 
\frac{1}{\alpha (\alpha^2 + \omega_0^2)} \,
\frac{2\omega_D^2-3\omega_D\alpha-9\alpha^2}{3\alpha + 2\omega_D}
\left\{ \frac{-\omega_D-6\alpha}{9(2\omega_D^2-3\omega_D\alpha-9\alpha^2)}-
\frac{1}{3\alpha+2\omega_D}+ \frac{1}{ 3 \alpha}
-\frac{2\alpha}{3(\omega_0^2+\alpha^2)}
\right\}\, , \notag \\
q_2^{(2)}(t) &=   e^{-(\omega_D/3+\alpha)t} \, \frac{2\omega_D-3\alpha}{2\omega_0^2}  \, 
\frac{1}{9\omega_D\alpha (\alpha^2 + \omega_0^2)} \,
\left\{ -3\omega_D+\omega_D^2\bigg(2t+\frac{2}{2\omega_D-3\alpha}+
\frac{2}{\alpha}- \frac{4\alpha}{\omega_0^2+\alpha^2}\bigg)
\right\}\, , \notag \\
q_2^{(3)}(t) &= e^{-\omega_D t/3}\frac{1}{9\omega_0^2 \alpha}    {\rm Re} \bigg\{
e^{i\omega_0 t } \frac{(2\omega_D+3i\omega_0)(-9\alpha^2-3\alpha\omega_D+2\omega_D^2+9\omega_0^2
-18i\omega_0\alpha-3i\omega_D\omega_0)}{2\omega_D+3\alpha+3i\omega_0}\, \notag \\
&\times \bigg[\frac{-\omega_D-6i\omega_0-6\alpha}{3(2\omega_0^2-\alpha^2-3i\alpha\omega_0)(-9\alpha^2-3\alpha \omega_D+2\omega_D^2+9\omega_0^2-18i\omega_0\alpha
-3i\omega_D\omega_0)}\, \notag \\ &-\frac{3}{(2\omega_0^2-\alpha^2-3i\alpha\omega_0)(2\omega_D+3\alpha+3i\omega_0)}+\frac{1}{(-\omega_0+i\alpha)^2(2i\omega_0+\alpha)}-
\frac{2}{\alpha(-2\omega_0+i\alpha)^2}\bigg ] \bigg\} \, , \notag \\
q_2^{(4)}(t) &= e^{-(\omega_D/3+\alpha) t }\frac{1}{\omega_0^2 \alpha}    {\rm Re} \bigg\{
e^{i\omega_0 t } \frac{3\alpha-2\omega_D-3i\omega_0}{2\omega_0^2+3i\omega_0\alpha-\alpha^2}\bigg [\frac{-2\omega_D-12i\omega_0}{6\omega_D+9i\omega_0}-\frac{2\omega_D^2-3i\omega_D \omega_0+9\omega_0^2}{3(2\omega_D+3i\omega_0)^2}\, \notag \\
&+\frac{(2\omega_D^2-3i\omega_D\omega_0+9\omega_0^2)t}{9(2\omega_D+3i\omega_0)}-\frac{2\omega_D^2-3i\omega_D\omega_0+9\omega_0^2}{3(2\omega_D+3i\omega_0)(-2\omega_D-3i\omega_0+3\alpha)}\, \notag\\
&+(2\omega_D^2-3i\omega_D\omega_0+9\omega_0^2)\bigg(
-\frac{1}{3(2\omega_D-3i\omega_0)^2}-
\frac{-2i\omega_0+\alpha}{9(2\omega_0^2+3i\alpha\omega_0-\alpha^2)(2\omega_D+3i\omega_0)}\, \notag \\
&-\frac{1}{9\alpha(2\omega_D+3i\omega_0)}+\frac{i\omega_0+\alpha}{9(2\omega_D+3i\omega_0)(2\omega_0^2+3i\alpha\omega_0-\alpha^2)}\bigg)\bigg]\bigg\}\, . \notag \\
\end{align}

-----------------